\documentclass[aps,prl,a4paper,noshowpacs,english,superscriptaddress,twocolumn]{revtex4}
\usepackage{amsmath,lipsum,epsfig,amssymb,amsfonts,mathrsfs,bm,dsfont}
\usepackage{amsthm,array}
\usepackage{amsmath}
\usepackage{CJK}
\usepackage{bm}
\usepackage{babel}
\usepackage{graphicx}
\allowdisplaybreaks
\usepackage{braket}
\usepackage[breaklinks]{hyperref}
\usepackage{color}
\usepackage{makecell}
\usepackage{tabularx}
\usepackage{multirow}
\usepackage{longtable}
\hypersetup{colorlinks=true, linkcolor=blue, citecolor=blue, filecolor=blue, urlcolor=blue}

\begin{document}

\title{Midgap states induced by Zeeman field and $p$ wave superconductor pairing}

\author{Yuanjun Jin}
\email[]{yuanjunjin@m.scnu.edu.cn}
 \affiliation {Guangdong Basic Research Center of Excellence for Structure and Fundamental Interactions of Matter, Guangdong Provincial Key Laboratory of Quantum Engineering and Quantum Materials, School of Physics, South China Normal University, Guangzhou 510006, China}
\affiliation{Division of Physics and Applied Physics, School of Physical and Mathematical Sciences, Nanyang Technological University, Singapore 637371, Singapore}
\author{XingYu Yue}
\affiliation{Physics Department and Guangdong-Hong Kong Joint Laboratory of Quantum Matter, the University of Hong Kong, Pokfulam Road, Hong Kong, China}

\author{Yong Xu}
\affiliation{Institute of Micro/Nano Materials and Devices, Ningbo University of Technology, Ningbo 315016, Zhejiang, China}

\author{Xiang-Long Yu}
\affiliation{Shenzhen Institute for Quantum Science and Engineering, Southern University of Science and Technology, Shenzhen 518055, China}
\affiliation{International Quantum Academy, Shenzhen 518048, China}

\author{Guoqing Chang}
\email[]{guoqing.chang@ntu.edu.sg}
\affiliation{Division of Physics and Applied Physics, School of Physical and Mathematical Sciences, Nanyang Technological University, Singapore 637371, Singapore}

\begin{abstract}
The one-dimensional Su-Schrieffer-Heeger (SSH) model is central to band topology in condensed matter physics, which allows us to understand and design distinct topological states. In this work, we find another mechanism to analogize the SSH model in a spinful system, realizing an obstructed atomic insulator by introducing intrinsic spin-orbit coupling and in-plane Zeeman field. In our model, the midgap states originate from a quantized hidden polarization with invariant index $\mathbb{Z}_2$ (0; 01) due to the local inversion symmetry breaking. When the global inversion symmetry is broken, a charge pumping is designed by tuning the polarization. Moreover, by introducing the $p+ip$ superconductor pairing potential, a new topological phase dubbed obstructed superconductor (OSC) is identified. This new state is characterized by invariant index $\mathbb{Z}_2$ (0; 01) and nonchiral midgap states. More interestingly, these nonchiral edge states result in a chiral-like nonlocal conductance, which is different from the traditional chiral topological superconductor. Our findings not only find another strategy to achieve a spinful SSH model but also predict the existence of OSC, providing a promising avenue for further exploration of its transport properties.
\end{abstract}



\maketitle

The one-dimensional (1D) Su-Schrieffer-Heeger (SSH) model, regarded as a prototypical example of topological insulators, is foundational to the field of band topology  \cite{ssh,sshe}. The key physics underlying the SSH model is the presence of alternating hopping integrals, resulting from the Peierls instability in the 1D spinless chain, which gives rise to quantized polarization and associated boundary states. This model provides an intuitive framework to understand the emergence of topological properties, such as topological invariants and bulk-boundary correspondence. By modifying the SSH model, numerous extended versions with different interactions have been extensively studied in the last decades \cite{sshinter,sshrmp}. For instance, by incorporating a staggered on-site potential, the SSH model evolved into the Rice-Mele model to investigate solitons \cite{ricemele,Berryphase}. Under a cyclic adiabatic evolution, the Rice-Mele model, playing as a charge pumping, provides perspective to the origin of nonzero Chern in quantum anomalous Hall insulators \cite{haldane1,Yu2010,Chang2013}. After that, topological states extend to $\mathbb{Z}_2$ topological insulators, topological superconductors (TSCs), Weyl semimetals, and Dirac semimetals, $etc$ \cite{grapheneTI1,Bernevig2006,Konig2007,m9,m10,m11,m12,m13,Wan2011,Weng2015,huang2015weyl,Xu2015,Lv2015,WangPhysRevB.85.195320,tci,rev1,rev2,rev3,rev5,rev4,chang2017,chang2018topological,jin2020,jindw}.

Recently, the extended two-dimensional (2D) SSH models have attracted considerable attention \cite{massh,m24,m27,nanol,reprl,reprr}, such as the nontrivial topological states with vanishing Berry Curvature \cite{reprl}. When a $\pi$-flux is applied to the 2D SSH models, it leads to the realization of quantized electric multipole insulators with corner states, sparking the exploration of high-order topological insulators (HOTIs) \cite{qemi,sciad,PhysRevLett.119.246402,ben2019,wieder2020strong,prbarxiv,tdai,nonnp,reprr,chan2023designer,y1,y2,y3,y4}. According to topological quantum chemistry (TQC) theory \cite{TQC,tqcnc}, symmetry indicators \cite{SI1,SI2,SI3,SI4}, and other theories \cite{ot1,ot2,ot3,ot4}, HOTIs can be identified by verifying whether the system is an obstructed atomic insulator (OAI), in which the valence electrons occupy empty Wyckoff positions in the lattice. Due to the misalignment of the obstructed Wannier charge center (WCC) with the occupied Wyckoff position, a clipped 2D OAI would exhibit midgap states. While TQC and other theoretical approaches provide powerful diagnostic tools for identifying OAIs, the SSH model remains essential for understanding the fundamental physical mechanisms behind OAIs \cite{oai1,oai2,oai3,oai4,oai5}.  The current SSH model is still limited to the alternating hopping integrals in spinless systems; other mechanisms to generate quantized polarization and bound states in spinful systems are rarely explored.

In this letter, we present a different strategy to achieve a spinful SSH model and to reveal the physical origin of the midgap states in OAI. In our model, the midgap states originate from quantized hidden polarization due to intrinsic spin-orbit coupling (SOC) and in-plane Zeeman field. Furthermore, by introducing the $p+ip$ superconductor pairing potential, one unique superconducting phase dubbed obstructed superconductor (OSC) is identified. Unlike the traditional chiral TSC, the OSC features nonchiral bound states but chiral-like nonlocal conductance due to the difference of normal electron tunneling (NET) when the electric field is reversed.

\begin{figure}
	\centering
	\includegraphics[scale=0.056]{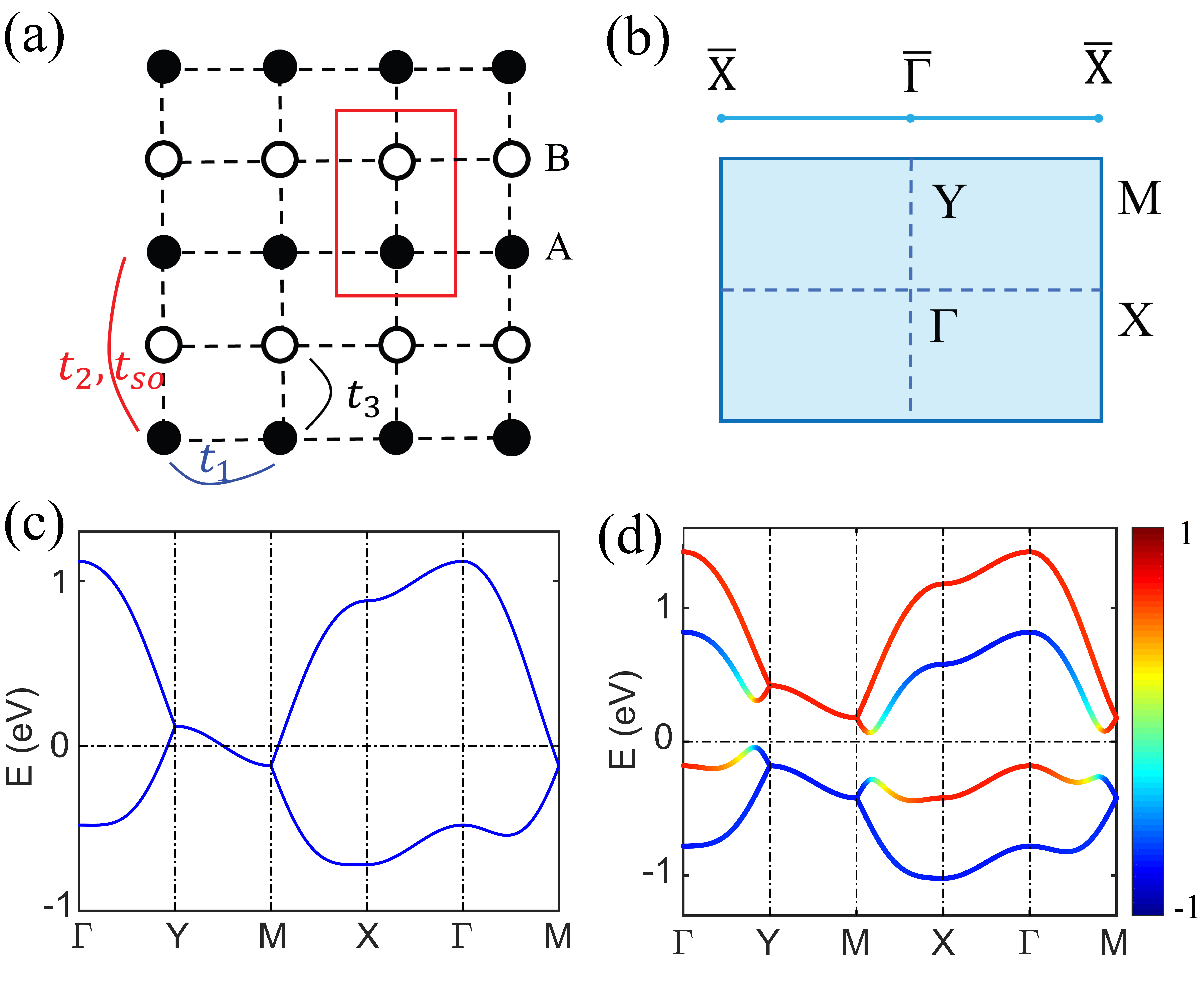}
	\caption{(a) 2D rectangle lattice includes two sites, A and B, with Wyckoff positions (0, 0.25) and (0, 0.75), respectively. The red rectangle represents the unit cell, and the curves are marked with the hopping parameters $t_i$ ($i$=1, 2, 3) and intrinsic SOC $t_{so}$. (b) The 2D BZ and the projected 1D BZ along $y$ axis. (c) Band structure in the presence of SOC. (d) Band structure with in-plane Zeeman field along $x$ direction. The color bar represents the spin direction along the $x$ axis.
		\label{figure-1}}
\end{figure}

\begin{figure}
	\centering
	\includegraphics[scale=0.059]{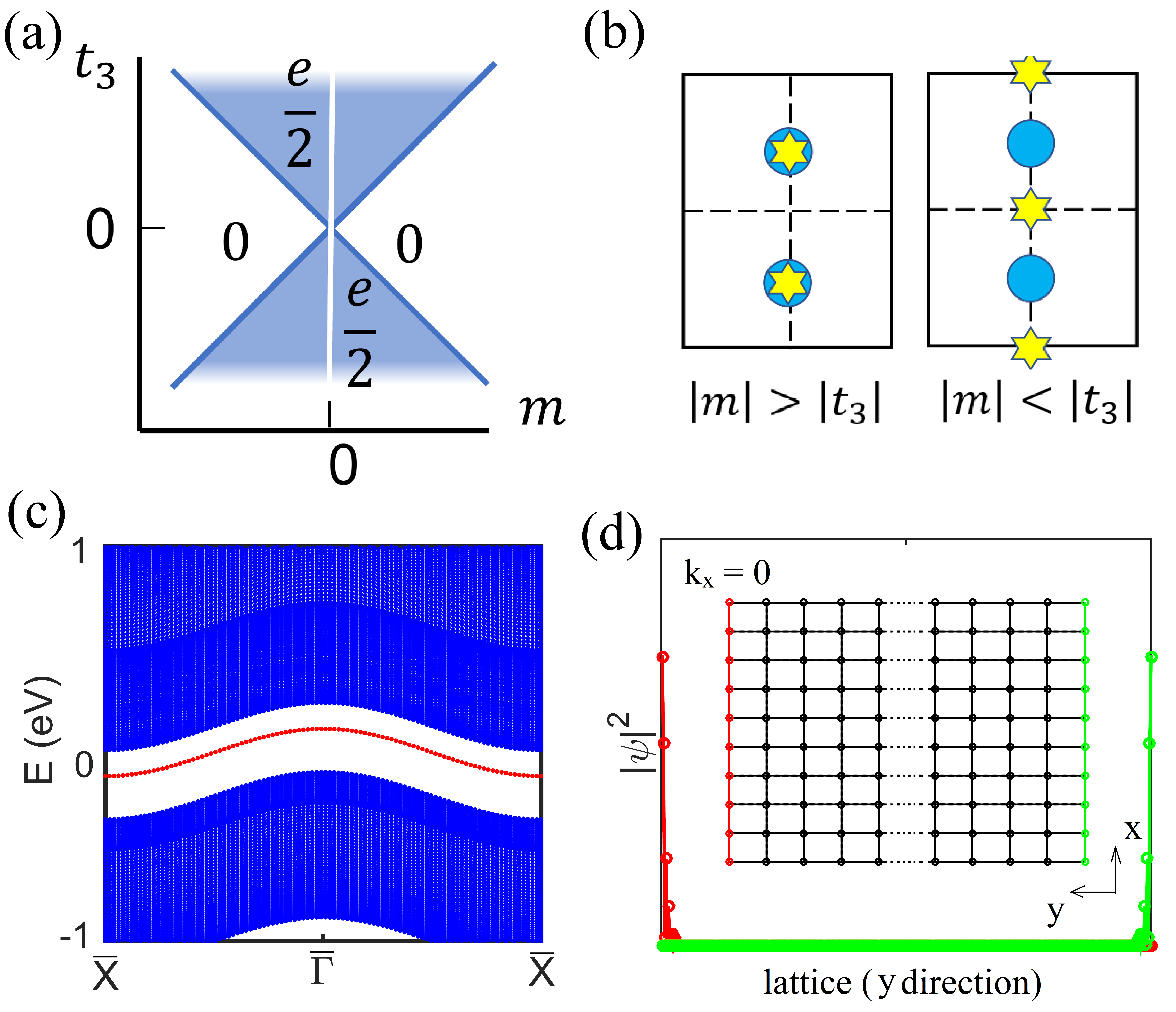}
	\caption{(a) Phase diagram depending on $t_3$ and $m$. (b) WCC and nuclei positions in the unit cell. Blue spheres denote the nuclei, and yellow stars present WCC.  (c) Edge states in the ribbon along $x$ axis. The width of this ribbon is about 100 unit cells. (d) The real-space probability distribution at $k_x$ = 0. Red and green indicate the states are localized on the two edges of the ribbon.
		\label{figure-2}}
\end{figure}

We construct a two-site tight-binding model for the inversion-symmetric layer group $Pmam$ (No. 40). As shown in Fig. \ref{figure-1}(a), there are two sites A and B as inversion-partners located at the Wyckoff positions (0, 0.25) and (0, 0.75), respectively. Figure \ref{figure-2}(b) displays the Brillouin zone (BZ) and its projection along the $y$-axis. In addition to the inversion symmetry ($\mathcal{I}$), the spatial symmetries include the glide mirror plane $\widetilde{\mathcal{M}}$$_x$=$\{\mathcal{M}_{x} | t \}$ and the screw axis $\widetilde{\mathcal{C}}$$_{2y}$=$\{\mathcal{C}_{2y} | t \}$, where ${\mathcal{M}_x}$ is reflection about \emph{yz} plane and ${\mathcal{C}_{2y}}$ is two-fold rotation along $\emph{y}$ axis, $t=(0,\frac{1}{2})$ is a half lattice translation along the $\emph{y}$ axis. Both $\widetilde{\mathcal{M}}$$_x$  and $\widetilde{\mathcal{C}}$$_{2y}$ switch sites A and B. Assuming each site has an $\emph{s}$-like orbit with two spin, the basis sets are  $\{\ket{\emph{A}}, \ket{\emph{B}}\} \otimes \{\ket{\uparrow}, \ket{\downarrow}\}$.  In the second quantization form, the tight-binding Hamiltonian is derived as
\begin{equation}\label{eqs1}
\begin{aligned}
H &= \frac{t_1}{2} \sum_{\left \langle ij \right \rangle}\sum_{\sigma}(a_{i,\sigma}^{\dag}a_{j,\sigma}+b_{i,\sigma}^{\dag}b_{j,\sigma})\\
&+ \frac{t_2}{2} \sum_{\left \langle \left \langle ij \right \rangle \right \rangle}\sum_{\sigma}(a_{i,\sigma}^{\dag}a_{j,\sigma}+b_{i,\sigma}^{\dag}b_{j,\sigma})\\
&+ \frac{t_3}{2}\sum_{\left \langle ij \right \rangle}\sum_{\sigma}(a_{i,\sigma}^{\dag}b_{j,\sigma}+b_{j,\sigma}^{\dag}a_{i,\sigma}) \\
&+\frac{it_{so}}{2} \sum_{\left \langle \left \langle ij \right \rangle \right \rangle}\sum_{\sigma\sigma^{\prime}}(a_{i,\sigma}^{\dag} s_{\sigma \sigma^{\prime}}^{z} a_{j,\sigma^{\prime}}-b_{i,\sigma}^{\dag} s_{\sigma \sigma^{\prime}}^{z} b_{j,\sigma^{\prime}}) \\
&+ m\sum_{i}\sum_{\sigma\sigma^{\prime}}(a_{i\sigma}^{\dag}\sigma_{\sigma \sigma^{\prime}}^{x} a_{i \sigma^{\prime}}+b_{i\sigma}^{\dag}\sigma_{\sigma \sigma^{\prime}}^{x} b_{i \sigma^{\prime}}),
\end{aligned}
\end{equation}
where $a_{i,\sigma}^{\dag}$ and $b_{i,\sigma}^{\dag}$ are electron creation operators at the sites A and B in unit cell $i$ with the spin $\sigma$, ${t}_{c}$ $(c=1, 2, 3)$ are hopping parameters and $t_{so}$  is the intrinsic SOC, and $m$ is the external Zeeman field strength along $x$ direction for simplicity, see the hoppings in Fig. \ref{figure-1} (a). In the following calculation, all the hopping parameters are provided in the Supplemental Materials (SM) \cite{SM}. In this centrosymmetric 2D lattice, the symmetry of the two sites is noncentrosymmetric because neither of them serves as an inversion center. Therefore, the system is centrosymmetric but locally noncentrosymmetric, giving rise to the intrinsic SOC term. In this case, since $s_z$ is conserved, the Hamiltonian can be decoupled into spin-up and spin-down sectors. Besides, the hidden polarization arises within a centrosymmetric system because the A-B sublattice with site symmetry group $C_{2v}$ individually breaks the $\mathcal{I}$ symmetry, creating a local dipole field compensated by its inversion counterpart \cite{hidden}.

Based on the above Hamiltonian Eq. (\ref{eqs1}), the electronic band dispersion is obtained, as shown in Figs. \ref{figure-1}(c). The energy bands show a fourfold degenerate nodal line along the Y-M path ($k_y$=$\pi$) in the presence of SOC. The nodal line is protected by the $\mathcal{IT}$ and $\widetilde{\mathcal{C}}$$_{2y}\mathcal{T}$ symmetries, see the symmetry arguments in the SM \cite{SM}. Next, we apply an in-plane Zeenman field to break the mirror plane $\mathcal{M}_z$ and $\mathcal{T}$ symmetries, to open a global gap. The Zeeman field is applied along the $x$ direction, which causes a mixing of the spin-up and spin-down states and lifts the energy degeneracy; see the spin-resolved energy bands in Fig. \ref{figure-1} (d). Since the global energy gap is open, we examine the topological polarization. We employ the homotopy description to systems with additional point group symmetries. Our four-band model with in-plane Zeeman field possesses $\mathcal{C}_{2z}\mathcal{T}$ symmetry so that one can identify the space of Hamiltonians as the coset space\cite{c2t,bou,bou20},
\begin{equation}
	M_{(2,2)}=O(4)/O(2) \times O(2),
\end{equation}
which is called real Grassmannian. The lowest non-trivial homotopy group is  $\pi_1(M_{(2,2)})=\mathbb{Z}_2$, see more details in the SM \cite{SM}.

Since $\mathcal{I}$ symmetry is conserved, the hidden polarization can be captured by the parity eigenvalues of the high symmetry invariant points for occupied bands using Eq. (S19). The corresponding parity eigenvalues for the two occupied bands are given in Table S1 in SM \cite{SM}. The hidden polarization phase diagram is shown in Fig. \ref{figure-2} (a). If the strength of the nearest hopping parameter $t_3$ is larger than the magnitude of the Zeenman field, that is $\left|t_3 \right| > \left|h \right|$, the polarization along the $y$ axis is quantized to $\frac{e}{2}$ except for $m$ = 0. If $\left|t_3 \right| < \left|h \right|$, the polarization in both directions vanishes. In addition, this state can also be characterized by the invariant index $\mathbb{Z}_2$ (0; 01), which describes the 1D polarization related to the geometry of the system (see more details in SM \cite{SM}).

The WCC calculation shows that in the nonpolarized phase, the WCC coincides in position with the nuclei [Fig. \ref{figure-2} (b), left panel]. In contrast, in the polarized phase, the WCC is symmetrically positioned in the middle of the neighboring A and B sites due to the $\mathcal{I}$ symmetry [Fig. S1 (a) in SM \cite{SM}], which refers to the obstructed WCC in TQC theory, as shown in the right panel in Fig. \ref{figure-2} (b) and Fig. S1 in SM \cite{SM}. This means the hidden polarization arises from the electronic rearrangement driven by the competition between the nearest hopping and the Zeeman field. Such hidden polarization can lead to interesting phenomena such as the emergence of midgap states, see Fig. \ref{figure-2} (c). The midgap states are doubly degenerate due to the $\mathcal{I}$ symmetry and are individually localized in two edges of the ribbon model, see  Fig. \ref{figure-2} (d). The presence of this hidden polarization has important consequences for the electronic and transport properties, including charge conduction and novel optical response.


We next consider a new SOC term to break $\mathcal{I}$ symmetry, which is given by
\begin{equation}\label{op111}
V=\frac{it_{so}^{\prime}}{2} \sum_{\left \langle ij \right \rangle}\sum_{\sigma\sigma^{\prime}}a_{i,\sigma}^{\dag} \sigma_{\sigma \sigma^{\prime}}^{z} b_{j,\sigma^{\prime}}. \\
\end{equation}
Once the $\mathcal{I}$ symmetry is broken, the WCC becomes asymmetric and deviates from the middle of the A and B sites, see Fig. S1 (b) in SM \cite{SM}. As a result, the previously hidden polarization becomes tunable. The polarization is obtained by below Eq. (\ref{Polar}),
\begin{equation}\label{Polar}
	P=\frac{e}{(2\pi)^2}\mathrm{Im}\sum_{n}\int_{0}^{2\pi}A_n(\textbf{k})d\textbf{k}
\end{equation}
where $A_{n}(\textbf{k})=i\left \langle u_{n\textbf{k}} \left| \nabla_{\textbf{k}}  u_{n\textbf{k}}\right.  \right \rangle$ is Berry connection. 

\begin{figure}
	\vspace{0.2cm}
	\centering
	\includegraphics[scale=0.059]{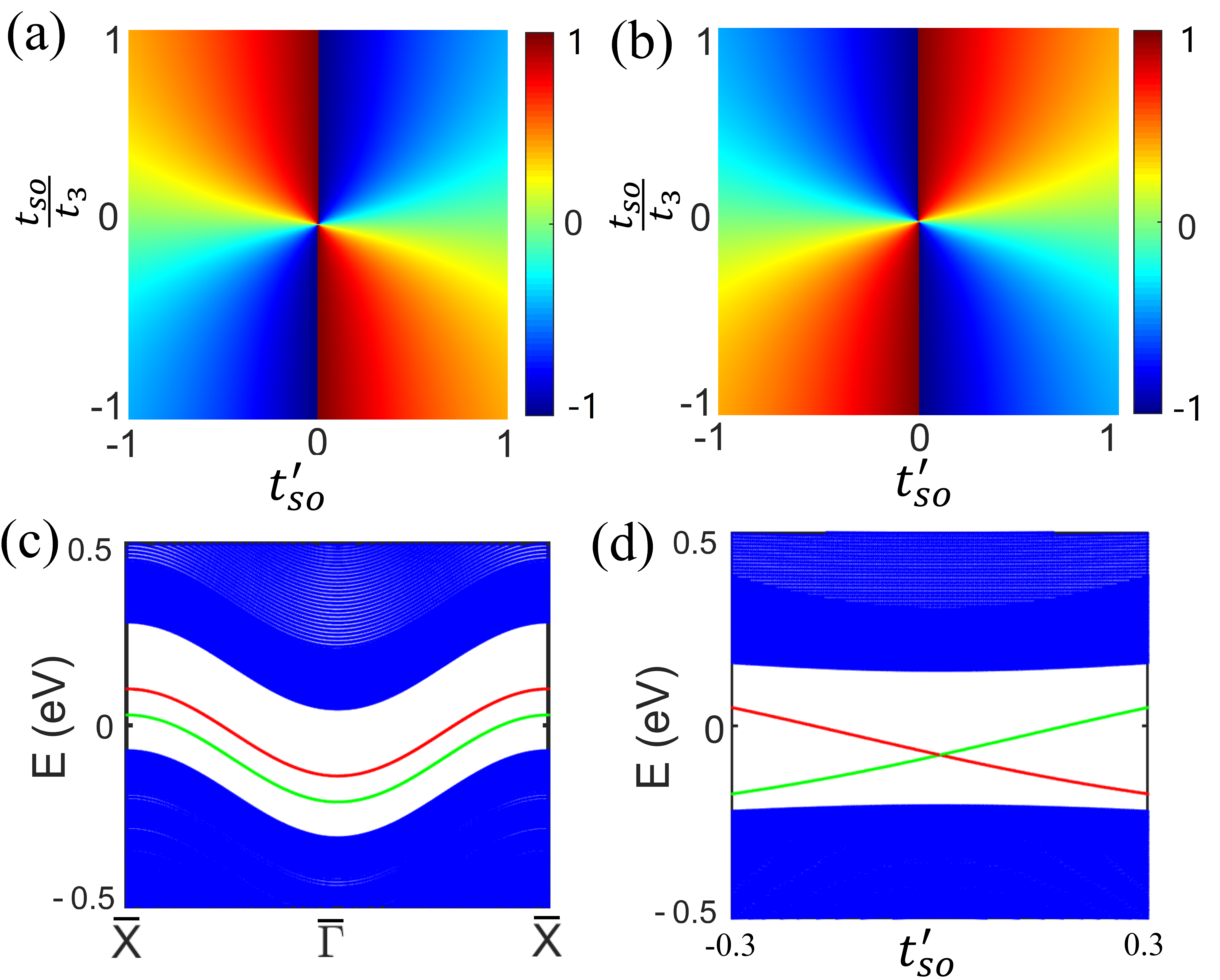}
	\caption{(a) Polarization as a function of $t_{so}^{\prime}$, $t_{so}$ and $t_3$ = 1. The unit is $2 \pi$. (b) Polarization as a function of $t_{so}^{\prime}$, $t_{so}$ and $t_3$ = $-1$. The unit is $2 \pi$. (c) The edge state splitting when $\mathcal{I}$ symmetry is broken with $t^{\prime}_{so}$=0.1. (d) Charge pumping as the revolution of $t^{\prime}_{so}$
		\label{figure-3}}
\end{figure}

The simulation shows that the magnitude and direction of polarization are locked to the parameter space of $t_3$, $t_{so}$, and $t_{so}^{\prime}$. Fig. \ref{figure-3} (a) and (b) show the calculated polarization for arbitrary $t_{so}$ and $t_{so}^{\prime}$, while $t_3$ is normalized to 1 for simplicity. When $t_{so}$ is zero, the polarization vanishes, indicating that the intrinsic SOC plays a critical role in the polarization. When intrinsic SOC is present and a finite $t_{so}^{\prime}$ changes sign, the polarization jumps between its maximum and minimum values. The change of the polarization leads to the lifting of the degeneracy of the edge states, as shown in Fig. \ref{figure-3} (c). Based on the above analysis, we design a charge pumping $H(t^{\prime}_{so},k_y)$ when $t^{\prime}_{so}$ is in a cyclic evolution. The corresponding edge states are shown in Fig. \ref{figure-3} (d). In such an adiabatic process, if $t_{so}^{\prime}$ changes in time through the zero point, a charge of $e$ is pumped across the insulator.

We next examine the topology under Cooper pairing in the $p+ip$  form for the spin-up sector, which is permitted in the $C_{2v}$ point group.
In the Bogliubov-de Genens (BdG) representation, the Hamiltonian in momentum space can be written as
\begin{equation}\label{ham1}
	H_{BdG}=\frac{1}{2}\sum_\mathbf{k}\Psi^{\dagger}_\mathbf{k\uparrow}
	\left ( \begin{matrix}
		H_{\mathbf{k}\uparrow}     & \mathbf{\Delta^\dagger_{\mathbf{k}\uparrow}}   \\
		\mathbf{\Delta_{\mathbf{k}\uparrow}}  & -H^*_{-\mathbf{k}\uparrow}
	\end{matrix} \right ) \Psi_{\mathbf{k}\uparrow}
\end{equation}
where $\Psi_{\mathbf{k}\uparrow}=(c_{\mathbf{k},A,\uparrow },c_{\mathbf{k},B,\uparrow },c_{-\mathbf{k},A,\uparrow }^{\dagger},c_{-\mathbf{k},B,\uparrow }^{\dagger})^\top$, $H_{\mathbf{k}\uparrow}$ is given by
\begin{widetext}
	\begin{equation}\label{hambdg}
		H_{k\uparrow}(k)=t_1 cosk_x  + t_2 cosk_x  + t_3 cos\frac{k_y}{2}\tau_x+t_{so}sink_y \tau_z.
	\end{equation}
\end{widetext}
Here,  $\mathbf{\Delta_{k\uparrow}} =\Delta_1  sin\frac{k_y}{2} \sigma_x+i\Delta_2 sink_x$ is the pairing order parameter, which shows the $p$ wave (spin-triplet) symmetry as $\mathbf{\Delta_k} = -\mathbf{\Delta_{-k}}$. $\Delta_1$  and $\Delta_2$  represent pairing order magnitudes along $y$ and $x$ directions, respectively.

\begin{figure}
	\centering
	\includegraphics[scale=0.057]{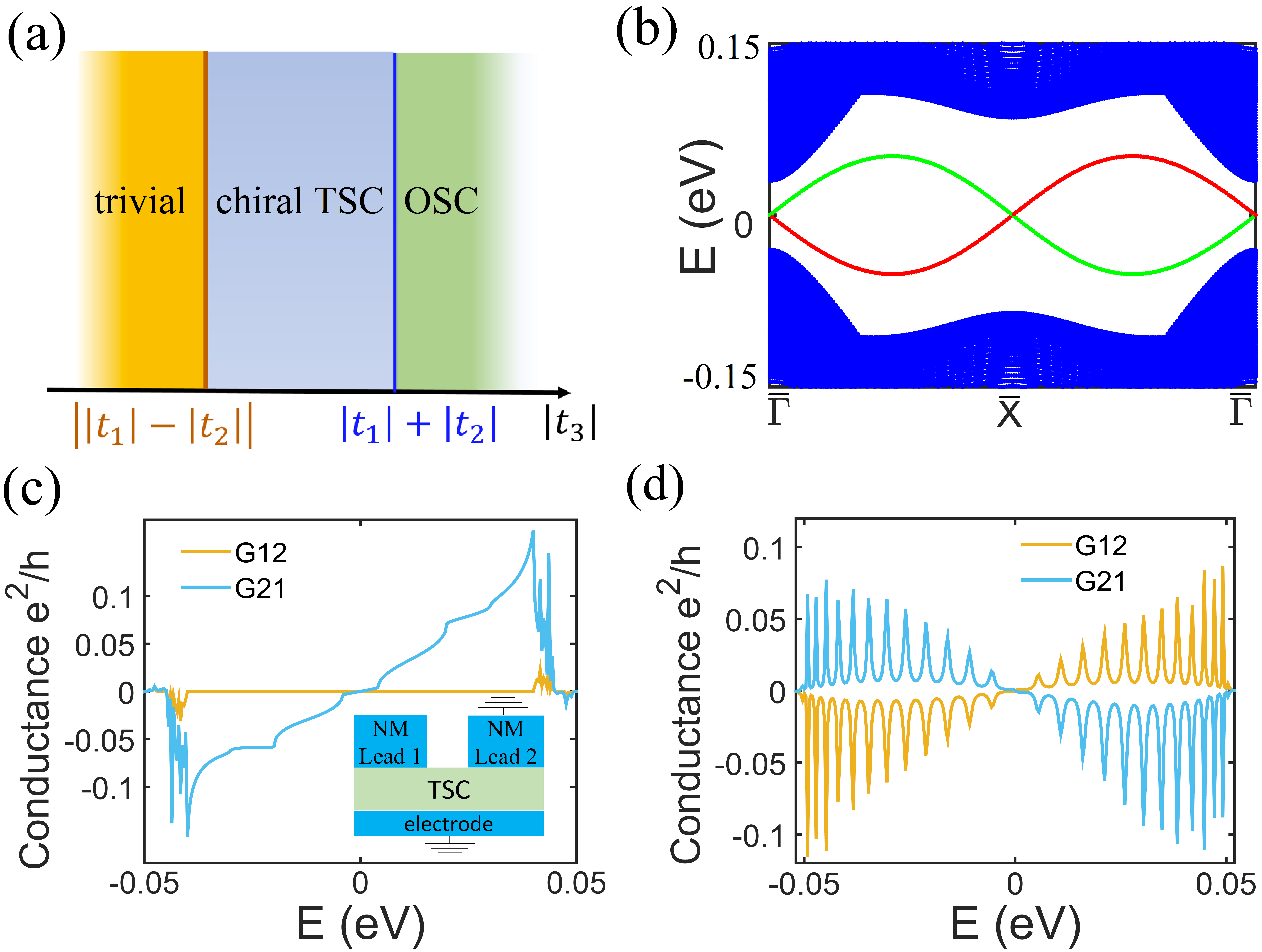}
	\caption{(a) Topological phase diagram (b) Edge states of obstructed SC, red and green indicate the states that come from two edges of the ribbon. (c) Nonlocal conductance of chiral TSC in our model. The insert shows the schematic three-terminal device with two normal metal leads. The figure corresponds to the setup for measuring $G_{12}$ and $G_{21}$. The bias voltage $V_1$ is applied to lead 1, while lead 2 and the superconductor are grounded. (d) The nonlocal conductance of OSC in our model.
		\label{figure-4}}
\end{figure}

It is easy to verify that $H_{BdG}$ respects the particle-hole symmetry $\mathcal{P} = s_x\kappa$ and the inversion symmetry $\mathcal{I} = s_z \tau_x$, where $s$ and $\tau$ are Pauli matrices in particle-hole and A-B sublattice spaces, respectively, and $\kappa$ is the complex conjugation. Thus, the band topology can be obtained by the parity eigenvalues at high symmetry invariant points, see details in SM \cite{SM}. Since $\mathcal{P}^2=+1$, the $H_{BdG}$ belongs to the D symmetry class. The topological index is characterized by $n_{1D}=\pi_1(O(4))=\mathbb{Z}_2$ and $n_{2D}=\pi_2(O(4))=\mathbb{Z}$, which denote the closed paths in $H_{BdG}$ manifold. To reveal the topological phase diagram, the pairing term $\mathbf{\Delta_k}$ can be considered as a Dirac Hamiltonian; the first three items in Eq. (\ref{hambdg}) are massive terms to tune the topological phases.  Three different phases, OSC, chiral TSC, and trivial SC, characterized by different invariants, are identified in this system. The phase diagram is shown in Fig. \ref{figure-4} (a), the horizontal line is $|t_3|$, while $|t_1|+|t_2|$ and $||t_1|-|t_2||$ are critical points, see details in SM \cite{SM}.  For generic $t_1$, $t_2$, and $t_3$, the condition reads
\begin{equation}
\begin{cases}
|t_1|+|t_2|<|t_3| & \text{OSC} \quad \mathbb{Z}_2=(0; 01)\\	
||t_1|-|t_2|| <|t_3|<|t_1|+|t_2| & \text{chiral TSC} \quad \mathbb{Z}=1 \\
|t_3|<||t_1|-|t_2|| & \text{trivial SC} \quad \mathbb{Z}_2=(0; 00).
\end{cases}
\end{equation}
Here, the bulk gap closes only at the boundaries between these three distinct phases. The chiral TSC with Chern number $\mathbb{Z}=1$ and the corresponding chiral edge states are shown in Fig. S2 in SM \cite{SM}. Analogous to the OAI in TQC theory, the obstructed WCC for a BdG Hamiltonian, implies the presence of OSC, see the obstructed WCC spectrum in Fig. S3 (a) in SM \cite{SM}, which is also consistent with the invariant index $\mathbb{Z}_2$ (0; 01) obtained by parity values in Table S3 in SM \cite{SM}. Due to the obstructed WCC, the corresponding edge states in the OSC, are detached from the bulk, as shown in Fig. \ref{figure-4} (b). The red and green lines indicate that the states are separately localized on the two edges of the ribbon. 

To explore the transport properties of these novel edge states, we perform the calculation of the nonlocal differential conductance using the $\mathrm{Kwant}$\cite{kwant}, see the schematic of the setup in Fig. \ref{figure-4} (c). This device comprises an obstructed $p$-wave superconductor and two normal metal (MN) leads. The nonlocal conductance is given by 
\begin{equation}\label{nonlocal}
	\begin{aligned}
	&G_{ab}(eV_b) =\frac{e^2}{h}[-R^{NET}_{ab}+R^{CAR}_{ab}]_{eV_b=E}, \\
	&R^{NET(CAR)}_{ab}=\sum_{n,m}|r^{ee(he)}_{ab}(n;m)|^2,
	\end{aligned}
\end{equation}
where $r^{ee}_{ab}(n;m)$ and $r^{he}_{ab}(n;m)$ are coefficients of NET and the crossed Andreev reflection (CAR), respectively. The index $n$ ($m$) denotes the outgoing (incoming) channel in the NM lead $a$ (lead $b$) with $a \neq b$.

For comparison, we plot the nonlocal conductance $G_{12}$ and $G_{21}$ for the chiral TSC phase in our model, which reveals the chirality-sensitive nonlocal conductance.  As shown in Fig. \ref{figure-4} (c),  only $G_{21}$ is non-zero, while $G_{12}$ is zero, aligning with the chiral non-local conductance previously reported \cite{m11}. When $t_3$ is increased to 0.25, the system enters the OSC phase, and the resulting non-local conductance is presented in Fig. \ref{figure-4}(d). In contrast to the chiral TSC phase, both $G_{12}$ and $G_{21}$ in the OSC phase have finite values but with opposite directions, with the orientation of the nonlocal conductance tied to the direction of the bias voltage.  It indeed exhibits the opposite nonlocal conductance, indicating the chiral-like motion of the edge states. We further pesent the spectra of $R^{NET}_{12(21)}$ and $R^{CAR}_{12(21)}$ in Fig. S4 in SM \cite{SM}. The $R^{CAR}_{12}$ has same values with $R^{CAR}_{21}$, while the $R^{NET}_{12}$ and $R^{NET}_{21}$ share different values, suggesting that the nonlocal chiral-like conductance comes from the difference of the NET probabilities when the voltage changes the sign.  For fixed energy, the states on the same edge provide two channels with opposite directions and opposite electron and hole components, as shown in Fig. S5 in SM \cite{SM}. As a result, the nonlocal conductance $G_{12}$ and $G_{21}$ exhibit finite values, but with opposite directions. Such chiral-like nonlocal conductance originating from the nonchiral edge states is different from the traditional chiral TSC systems and can be considered as the signature of the  OSC detection.

In this work, we provide a new strategy to analogize the SSH model in a spinful system to reveal the physical origin of midgap states in OAI. Such midgap states protected by the hidden polarization stem from the intrinsic SOC due to the local $\mathcal{I}$ symmetry breaking. When the global $\mathcal{I}$ symmetry is broken, charge pumping is achieved through variations in polarization. Furthermore, by introducing $p+ip$ superconductor pairing, three different phases are identified. In addition to the traditional chiral TSC, we find a new phase of OSC, which is characterized by a nonzero invariant index $\mathbb{Z}_2$ (0; 01) and obstructed WCC. The chiral-like nonlocal conductance is obtained as the signature of the OSC, making it possible for experiments to detect it. Our results not only realize a spinful SSH model but also predict the OSC, providing a new platform to explore its transport properties.

~~~\\
~~~\\
The authors thank  M. H. Xie, J. F. Liu, P. H. Fu, and Z. J. Chen for helpful discussions. Work at Nanyang Technological University was supported by the National Research Foundation, Singapore, under its Fellowship Award (NRF-NRFF13-2021-0010), the Agency for Science, Technology and Research (A*STAR) under its Manufacturing, Trade and Connectivity (MTC) Individual Research Grant (IRG) (Grant No.: M23M6c0100), Singapore Ministry of Education (MOE) AcRF Tier 2 grant (MOE-T2EP50222-0014) and the Nanyang Assistant Professorship grant (NTU-SUG). Y. Xu was supported by the Scientific Research Starting Foundation of Ningbo University of Technology (Grant No. 2022KQ51), China Postdoctoral Science Foundation (Grant No. 2023M743783). X. L. Yu was supported by the Natural Science Foundation of Guangdong Province (Grant No. 2023A1515011852). Y. J. Jin was supported by the startup funding from South China Normal University (Grant No. 8S078628).\\
~~~\\

Yuanjun Jin and Xingyu Yue equally contributed to this work.
%
%
%

\end{document}